# BERT-based Ranking for Biomedical Entity Normalization

Zongcheng Ji, PhD[1], Qiang Wei, MS[1], Hua Xu, PhD[1]
[1]School of Biomedical Informatics, The University of Texas Health Science Center at Houston, Houston, TX, USA

**Abstract**

*Developing high-performance entity normalization algorithms that can alleviate the term variation problem is of great interest to the biomedical community. Although deep learning-based methods have been successfully applied to biomedical entity normalization, they often depend on traditional context-independent word embeddings. Bidirectional Encoder Representations from Transformers (BERT), BERT for Biomedical Text Mining (BioBERT) and BERT for Clinical Text Mining (ClinicalBERT) were recently introduced to pre-train contextualized word representation models using bidirectional Transformers, advancing the state-of-the-art for many natural language processing tasks. In this study, we proposed an entity normalization architecture by fine-tuning the pre-trained BERT / BioBERT / ClinicalBERT models and conducted extensive experiments to evaluate the effectiveness of the pre-trained models for biomedical entity normalization using three different types of datasets. Our experimental results show that the best fine-tuned models consistently outperformed previous methods and advanced the state-of-the-art for biomedical entity normalization, with up to 1.17% increase in accuracy.*

**Introduction**

Entity linking, which aims to link entity mentions detected in a document to their corresponding concepts in a given knowledge base (KB) or an ontology[1], is one of the fundamental tasks in information extraction. The main challenges of this task are (1) *ambiguity* – the same entity mention may be linked to multiple concepts, (2) *variation* – the same concept can be linked by different entity mentions, and (3) *absence* – entity mentions may not be linked to any concept in the given KB. In the biomedical domain, this task is also known as entity normalization or encoding. Unlike in the general domain where ambiguity is the primary challenge, variation is much more common than ambiguity in the biomedical domain[2,3]. Therefore, developing high-performance entity normalization algorithms that can alleviate the variation problem is of great interest to the biomedical community.

Many studies have focused on solving the variation challenge in the biomedical domain, resulting in development of rule-based methods[3–5], machine learning-based methods[6,7], and deep learning-based methods[2,8]. Kang et al.[5] developed a rule-based natural language processing (NLP) module containing 5 types of rules, to improve disease normalization in biomedical text. Ghiasvand and Kate[4] first automatically learned 554 edit distance patterns of term variations between all the synonyms of disorder concepts in the Unified Medical Language System (UMLS)[9] as well as between the entity mentions in the training data and their corresponding concepts in the UMLS. They then normalized the entity mentions in the test data by performing exact match between the variations generated by the learned patterns and an entity mention in the training data or a concept name in the given KB. Their system named UWM was the best system for the disease and disorder mention normalization task of the SemEval 2014 challenge[4,10]. D'Souza and Ng[3] proposed a multi-pass sieve system by defining 10 types of rules at different priority levels to measure morphological similarity between entity mentions and candidate concepts in the given KB. Leaman et al.[7] proposed a pairwise learning-to-rank method by adopting vector space model to represent entity mentions and concepts, and using a similarity matrix to measure the similarities between entity mentions and candidate concepts. Xu et al.[6] also proposed a pairwise learning-to-rank method by defining 3 kinds of features and employing the linear RankSVM[11] to normalize each positive adverse reaction mention to an entry in MedDRA. Their system achieved the best performance in the TAC 2017 ADR challenge[12]. Li et al.[2] proposed a convolutional neural network (CNN) architecture that regarded biomedical entity normalization as a ranking problem, which takes advantage of CNN in modeling semantic similarities between entity mentions and candidate concepts. The method outperformed traditional rule-based methods, achieving the state-of-the-art performance. Luo et al.[13] proposed a multi-view CNN with multi-task shared structure to normalize diagnostic and procedure names simultaneously in Chinese discharge summaries to standard concepts.

Although deep learning-based methods[2,13] have been successfully applied to biomedical entity normalization, they required pre-trained word embeddings that were often learned from a large corpus of unannotated texts. Word2vec[14] has been widely adopted to pre-train word embeddings from large corpora and was also used in the work of Li et al.[2] and Luo et al.[13]. Recently, ELMo[15] generalized traditional word embeddings to contextual word embeddings and advanced the state-of-the-art for several major NLP benchmarks when integrating contextual word embeddings with existing task-specific architectures. The Generative Pre-trained Transformer (GPT)[16] introduced minimal task-specific parameters and could be trained on the downstream tasks by simply fine-tuning the pre-trained parameters. Unlike ELMo and GPT, which used unidirectional language models for pre-training, Bidirectional Encoder Representations from Transformers (BERT) introduced masked language models to enable pre-training deep bidirectional representations and advanced the state-of-the-art for eleven NLP tasks[17]. Based on the BERT architecture, BioBERT[18] (BERT for Biomedical Text Mining) and ClinicalBERT[19–21] (BERT for Clinical Text Mining), which were domain-specific language representation models pre-trained on large-scale biomedical articles and clinical notes, were introduced to advance the state-of-the-art performance on many biomedical and clinical NLP tasks.

Despite promising work on the pre-trained BERT / BioBERT / ClinicalBERT models for many NLP tasks such as named entity recognition (NER), relation classification (RC) and question answering (QA) in both the general domain[17] and biomedical domain[18–22], no existing work has investigated the models for biomedical entity normalization. This task is very different from the above NLP tasks in that NER and QA are token-level tagging tasks and RC is single sentence classification task while biomedical entity normalization can be seen as sentence pair classification task, where we decide whether a candidate concept can be linked by a given entity mention. As a preliminary study, here we proposed an entity normalization architecture by fine-tuning the pre-trained BERT / BioBERT / ClinicalBERT models and conducted extensive experiments to evaluate the effectiveness of the pre-trained models for the entity normalization task using three different types of datasets in the biomedical domain.

*Table 1: Statistics of the three types of datasets used in this study.*

|  | **ShARe/CLEF** (Clinical Notes) | | **NCBI** (PubMed Abstracts) | | **TAC2017ADR** (Drug Labels) | |
| --- | --- | --- | --- | --- | --- | --- |
|  | train | test | train | test | train | test |
| #documents | 199 | 99 | 692 | 100 | 101 | 99 |
| #mentions | 5,816 | 5,351 | 5,921 | 960 | 7,038 | 6,343 |
| #mentions that are linkable | 4,175 | 3,601 | 5,921 | 960 | 6,991 | 6,325 |
| #mentions that are unlinkable | 1,641 | 1,750 | 0 | 0 | 47 | 18 |
| #concepts | 88,150 | | 9,664 | | 23,668 | |

**Methods**

*Datasets*

We used three different types of datasets in this study, namely ShARe/CLEF - the ShARe/CLEF eHealth 2013 Challenge corpus[23], NCBI - the NCBI disease corpus[24], and TAC2017ADR - the TAC 2017 ADR corpus[12]. Table 1 shows the statistics of the three datasets.

**ShARe/CLEF**: This dataset contains 298 de-identified clinical notes collected from a US intensive care data repository including discharge summaries, electrocardiograms, echocardiograms, and radiology reports, which was partitioned into 199 notes for training and development and 99 notes for testing. Based on a pre-defined annotation guideline, a disorder mention in each clinical note was manually annotated with its mapping concept unique identifier (CUI) within the SNOMED-CT subset of the UMLS[9]. If there was no mapping concept for a disorder mention, a CUI-less label (i.e., unlinkable) was assigned. We followed the guideline to construct the SNOMED-CT subset from the UMLS 2012AB, which contains 88,150 disorder concepts. Table 1 shows that 28.2% of the training mentions and 32.7% of the testing mentions were unlinkable, which illustrates the *absence* challenge of entity normalization.

**NCBI**: This dataset contains 792 PubMed abstracts, which was split into 692 abstracts for training and development, and 100 abstracts for testing. A disorder mention in each PubMed abstract was manually annotated with its mapping concept identifier in the MEDIC lexicon[25]. In this study, we used the July 6, 2012 version of MEDIC, which contains 7,827 MeSH identifiers and 4,004 OMIM identifiers, grouped into 9,664 disease concepts. Different from the ShARe/CLEF dataset, only those disorder mentions that can be mapped to a concept in MEDIC were annotated in NCBI. As a result, all the annotated disorder mentions have their corresponding concept identifiers.

**TAC2017ADR**: This dataset contains 200 drug labels, which was split into 101 labels for training and development, and 99 labels for testing. An adverse reaction in each drug label was manually annotated with its mapping MedDRA

Lower Level Term (LLT) and the corresponding Preferred Term (PT). If there was no ideal PT mapped for an adverse reaction mention, a High Level Term (HLT) or a High Level Group Term (HLGT) was provided if appropriate, otherwise an "unmapped" tag (i.e., unlinkable) was assigned to the mention. In this study, we constructed a KB from MedDRA v18.1, which contains 21,612 PTs, 1,721 HLTs, and 335 HLGTs, grouped into 23,668 unique concepts. Note that only 0.7% of the training mentions and 0.3% of the testing mentions were unlinkable in this dataset.

*Entity Normalization - Problem Definition*

Given an entity mention $m$ recognized from a sentence $x$ within a document $d$, and a KB which consists of a set of concepts, the task of entity normalization is to link $m$ to the corresponding concept $c$ in KB, $m \rightarrow c$. If there is no mapping concept in KB for $m$, then $m \rightarrow NIL$, where $NIL$ denotes that $m$ is unlinkable.

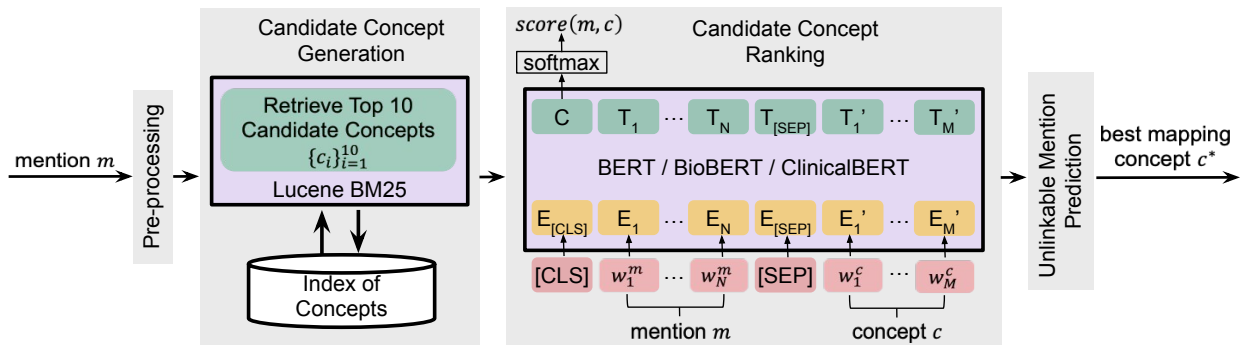

*Figure 1: System architecture for entity normalization used in this study.*

*Entity Normalization – System Architecture*

Figure 1 shows the system architecture for entity normalization used in this study, which consists of four modules: preprocessing, candidate concept generation, candidate concept ranking and unlinkable mention prediction.

- **Preprocessing**: We preprocessed each mention and each concept in KB with the following strategies.
  - *Spelling Correction* – For each mention in the ShARe/CLEF and NCBI datasets, we replaced all the misspelled words using a spelling check list as in previous work[2,3]. (e.g., fist → first, sytem → system, etc.)
  - *Abbreviation Resolution* – We used Ab3p[26] toolkit to detect the abbreviations within each document, and then replaced each mention in short-form abbreviation with its corresponding long form. (e.g., WT → Wilms tumor) Specifically, for the ShARe/CLEF and NCBI datasets, we also expanded all possible abbreviated disorder mentions using Schwartz and Hearst's algorithm[27] and a list of disorder abbreviations collected from Wikipedia as in previous work[2,3].
  - *Numeric Synonyms Resolution* – We replaced all the numerical words in the mentions and concepts to their corresponding Arabic numerals as in previous work[2,3,7]. (e.g., one / first / i / single → 1)
  - *Other Preprocessing* – Finally, we tokenized all the mentions and concepts by whitespace, removed all the punctuations, stemmed the tokens with the Porter stemmer and converted all the tokens into lower case ASCII. All of these were implemented using the CLAMP[28] toolkit.
- **Candidate Concept Generation**: We generated candidate concepts for each mention with the commonly used information retrieval (IR) based method[6,29–31], which included the following two steps. We first indexed all the concept names and training mentions with their concept ids. Then, we employed the traditional IR model of BM25[32] provided by Lucene to retrieve the top 10 candidate concepts $\{c_i\}_{i=1}^{10}$ for each mention $m$.
- **Candidate Concept Ranking**: We reranked the candidate concepts by fine-tuning the pre-trained BERT / BioBERT / ClinicalBERT models, where we transformed the ranking task as a sentence-pair classification task. Specifically, for each mention $m$ and a candidate concept $c$, we constructed a sequence [CLS] $m$ [SEP] $c$ as the input of the fine-tuning procedure, where [CLS] was the special word used for the classification output, and [SEP] was the special word used for separating $m$ and $c$. The output of the fine-tuning procedure was the final hidden state of the first word [CLS] of the input sequence, which was a fixed-dimensional word embedding $C \in \mathbb{R}^H$. The only new parameters added during the fine-tuning procedure were $W \in \mathbb{R}^{K \times H}$, which was used for the final classifier layer. Here $K = 2$ was the number of classifier labels. If $c$ is the mapping concept for $m$, the classifier label is 1, otherwise the label is 0. The probability of label=1 was computed with a softmax function, which was used as the ranking score of each pair $(m, c)$: $score(m, c) = P(label = 1|m, c) = softmax(CW^T)$.

- **Unlinkable Mention Prediction**: Because some entity mentions may not have any mapping concepts in KB, it is necessary to predict unlinkable mentions. If there were no candidate concepts returned from Lucene BM25, we predicted $m$ as an unlinkable mention and return $m \to NIL$ undoubtedly. Otherwise, we chose the top ranking concept $c^* = \arg\max_{c' \in \{c_i\}_{i=1}^{10}} score(m, c')$. Here, we validated whether $m \to c^*$ holds by adopting a simple and widely used method to learn a NIL-threshold $\tau$. Namely, if $score(m, c^*) > \tau$, then $m \to c^*$, otherwise $m \to NIL$. We learned the threshold $\tau$ from the training data with a small held-out development set.

*BERT Models*

In this study, we used the pre-trained BERT[33], BioBERT[34], and ClinicalBERT[19] models for the fine-tuning procedure. BERT models were trained on Wikipedia and BooksCorpus. BioBERT models were initialized with BERT$_{Base\_Cased}$ model and pre-trained with additional biomedical corpus including PubMed abstracts (PubMed), PubMed Central full-text articles (PMC), or PubMed+PMC. There were three types of publicly available ClinicalBERT[19–21] models trained with clinical notes from MIMIC-III (Medical Information Mart for Intensive Care III) critical care database[35]. Huang et al.[20] pre-trained the ClinicalBERT model from scratch with randomly sampled 100,000 clinical notes from MIMIC-III. Si et al.[19] pre-trained two ClinicalBERT models initialized from BERT$_{Base\_Cased}$ and BERT$_{Large\_Cased}$ with all the clinical notes from MIMIC-III. Alsentzer et al.[21] pre-trained two ClinicalBERT models initialized from BioBERT with all the clinical notes and all the discharge summaries from MIMIC-III. In this study, we investigated the two ClinicalBERT models at 300K training steps released by Si et al.[19]. More specifically, we investigated four different versions of BERT models (i.e., BERT$_{Base\_Cased}$, BERT$_{Base\_Uncased}$, BERT$_{Large\_Cased}$, BERT$_{Large\_Uncased}$), three different versions of BioBERT models (i.e., BioBERT$_{Base\_Cased+PubMed}$, BioBERT$_{Base\_Cased+PMC}$, BioBERT$_{Base\_Cased+PubMed+PMC}$), and two different versions of ClinicalBERT models (i.e., ClinicalBERT$_{Base\_Cased+MIMIC}$, ClinicalBERT$_{Large\_Cased+MIMIC}$).

*Parameters Settings*

For fine-tuning, most model hyperparameters were the same as those saved in the pre-trained model, with the exception of the batch size, learning rate, and number of training epochs[33]. In this study, we fixed the learning rate at 2e-5, tuned the batch size with 16 and 32, tuned the number of training epochs from 1 to 10, and saved the model with the best performance.

*Evaluation Metrics*

Following previous work[2,3], we evaluated the performance of different entity normalization algorithms in terms of accuracy, which was the percentage of entity mentions that were correctly normalized.

Table 2: Comparisons of different pre-trained models. The bold score denotes the best performance of each dataset.

|  | ShARe/CLEF | NCBI | TAC2017ADR |
| --- | --- | --- | --- |
| BM25 | 85.14 | 88.23 | 91.09 |
| BERT$_{Base\_Cased}$ | 90.62 | 88.85 | 92.62 |
| BERT$_{Base\_Uncased}$ | 90.58 | 88.65 | 92.97 |
| BERT$_{Large\_Cased}$ | 90.73 | 88.85 | 92.87 |
| BERT$_{Large\_Uncased}$ | 90.66 | 88.13 | 92.87 |
| BioBERT$_{Base\_Cased+PubMed}$ | **91.10** | 88.23 | **93.22** |
| BioBERT$_{Base\_Cased+PMC}$ | 90.99 | 88.65 | 92.97 |
| BioBERT$_{Base\_Cased+PubMed+PMC}$ | 91.09 | **89.06** | 93.17 |
| ClinicalBERT$_{Base\_Cased+MIMIC}$ | 90.62 | 88.96 | 92.70 |
| ClinicalBERT$_{Large\_Cased+MIMIC}$ | 90.88 | 88.13 | 92.94 |

**Results**

*Comparisons of different pre-trained models*

Table 2 shows the performance comparisons of different pre-trained models with the BM25 baseline for biomedical entity normalization. From the table, we see that (1) All the BERT / BioBERT / ClinicalBERT models outperformed the BM25 model by at least 5.44% (90.58 vs. 85.14) and 1.53% (92.62 vs. 91.09) on both the ShARe/CLEF and TAC2017ADR datasets. Most of them outperformed the BM25 model for the NCBI dataset by up to 0.83% (89.06 vs. 88.23) except $BERT_{Large\_Uncased}$, $BioBERT_{Base\_Cased+PubMed}$ and $ClinicalBERT_{Large\_Cased+MIMIC}$. (2) The BERT models with cased version were better than that with uncased version in most cases for biomedical entity normalization. (3) For the ShARe/CLEF and TAC2017ADR datasets, all the three BioBERT models outperformed the $BERT_{Base\_Cased}$ model and both the two ClinicalBERT models outperformed the corresponding $BERT_{Base\_Cased}$ and $BERT_{Large\_Cased}$ models. However, for the NCBI dataset, only $BioBERT_{Base\_Cased+PubMed+PMC}$ and $ClinicalBERT_{Base\_Cased+MIMIC}$ were better than $BERT_{Base\_Cased}$. (4) $BioBERT_{Base\_Cased+PubMed}$ achieved the best performance on both the ShARe/CLEF and TAC2017ADR datasets, while $BioBERT_{Base\_Cased+PubMed+PMC}$ achieved the best performance for the NCBI dataset.

*Comparisons with existing work*

We compared the following state-of-the-art methods with our best fine-tuned BERT-based ranking model.

- UWM[4]: the best challenge system on the ShARe/CLEF dataset, which is a rule-based system.
- TaggerOne[36]: the best machine learning-based system up to date on the NCBI dataset. It performs named entity recognition and normalization jointly, which is significantly different from our problem definition.
- Xu et al.'system[6]: the best challenge system on the TAC2017ADR dataset, which is a machine learning-based system.
- D'Souza & Ng's system[3]: the best rule-based system up to date on both the ShARe/CLEF and NCBI datasets.
- CNN-based ranking[2]: the best deep learning-based system up to date on both the ShARe/CLEF and NCBI datasets. Since we cannot completely reconstructed the KBs as used but not released in Li et al.'s work[2], we reimplemented the system and used the same settings as described in their paper. In addition, we employed word2vec[14] to train the word embeddings with a dimension size of 50 from all the clinical notes in MIMIC-III[19], the PubMed biomedical abstracts as used in Li et al.'s work[2], and the drug labels as used in Xu et al.'s work[6] for the ShARe/CLEF, NCBI, and TAC2017ADR datasets, respectively.

Table 3: Comparisons with existing work. The bold score denotes the best performance of each dataset.

|  | ShARe/CLEF | NCBI | TAC2017ADR |
|---|---|---|---|
| UWM[4] | 89.50 | NA | NA |
| TaggerOne[36] | NA | 88.80 | NA |
| Xu et al.'s system[6] | NA | NA | 92.05 |
| D'Souza & Ng's system[3] | 90.75 | 84.65 | NA |
| CNN-based ranking[2] | 90.30 | 86.10 | NA |
| CNN-based ranking (reimplement) | 88.97 | 86.67 | 90.24 |
| Our best BERT-based ranking | **91.10** | **89.06** | **93.22** |

Table 3 shows the performance comparisons of the state-of-the-art methods with our best fine-tuned BERT-based ranking models for biomedical entity normalization. The table shows that our best BERT-based ranking models consistently outperformed previous methods and achieved the state-of-the-art performance in terms of accuracy by 0.35%, 0.26% and 1.17% on the ShARe/CLEF, NCBI, TAC2017ADR datasets, respectively. Note that, due to we used different KBs, the results of our reimplemented CNN-based ranking on the ShARe/CLEF and NCBI datasets were different from that reported in Li et al.'s work[2].

*The impact of different batch sizes*

Table 4 shows the impact of different batch sizes on the three datasets. We compared batch sizes of 16 and 32 as suggested by Devlin et al.[17]. From the table, we observe that (1) For the NCBI and TAC2017ADR datasets, setting batch size as 16 achieved better performance than as 32. For the ShARe/CLEF dataset, there was no obvious difference between different batch size settings. (2) The best performance was achieved when batch size was set as 16 on all the three datasets.

Table 4: The impact of different batch sizes. The underlined score denotes that the performance of the model with the current batch size was better than the other choice. The bold score denotes the best performance of each dataset.

|  | ShARe/CLEF | | NCBI | | TAC2017ADR | |
| --- | --- | --- | --- | --- | --- | --- |
| batch size | 16 | 32 | 16 | 32 | 16 | 32 |
| BERT$_{Base\_Cased}$ | 90.56 | 90.62 | 88.85 | 88.65 | 92.62 | 92.56 |
| BERT$_{Base\_Uncased}$ | 90.56 | 90.58 | 88.65 | 88.13 | 92.97 | 92.65 |
| BERT$_{Large\_Cased}$ | 90.73 | 90.71 | 88.85 | 88.33 | 92.42 | 92.87 |
| BERT$_{Large\_Uncased}$ | 90.66 | 90.66 | 88.13 | 88.13 | 92.87 | 92.70 |
| BioBERT$_{Base\_Cased+PubMed}$ | **91.10** | 91.01 | 88.23 | 88.02 | **93.22** | 92.98 |
| BioBERT$_{Base\_Cased+PMC}$ | 90.81 | 90.99 | 88.65 | 88.65 | 92.97 | 92.89 |
| BioBERT$_{Base\_Cased+PubMed+PMC}$ | 91.01 | 91.09 | **89.06** | 88.85 | 93.17 | 92.89 |
| ClinicalBERT$_{Base\_Cased+MIMIC}$ | 90.62 | 90.54 | 88.96 | 88.44 | 92.70 | 92.67 |
| ClinicalBERT$_{Large\_Cased+MIMIC}$ | 90.88 | 90.73 | 88.13 | 88.02 | 92.94 | 92.80 |

**Discussion**

In this study, we developed an entity normalization architecture by fine-tuning the pre-trained BERT / BioBERT / ClinicalBERT models and conducted extensive experiments to evaluate the effectiveness of the pre-trained models for the entity normalization task using biomedical datasets of three different types. Our best fine-tuned models consistently outperformed previous methods and advanced the state-of-the-art on biomedical entity normalization by up to 1.17% increase in accuracy. To the best of our knowledge, this is the first study to apply and evaluate the pre-trained BERT / BioBERT / ClinicalBERT models for biomedical entity normalization.

From Table 2, we notice that although all the best fine-tuned models outperformed BM25 on the three datasets, it did not improve too much on the NCBI dataset (i.e., by up to 0.83%). BERT$_{Large\_Uncased}$ and ClinicalBERT$_{Large\_Cased+MIMIC}$ performed even worse than BM25. This indicates the difficulty of this dataset. Choosing an appropriate pre-trained model for this dataset is necessary. In the future, we will further investigate better methods for this dataset, e.g., tuning different learning rates to find a better fine-tuned model.

The BERT models with cased version were better than that with uncased version in most cases for biomedical entity normalization. This indicates that the BERT models with cased version could capture more precise contextualized word representations than that with uncased version, and they are benefit for the entity normalization task.

The three BioBERT models were initialized with BERT$_{Base\_Cased}$ and pre-trained with biomedical corpora[34]. The two ClinicalBERT models were initialized with BERT$_{Base\_Cased}$ and BERT$_{Large\_Cased}$, and pre-trained with clinical notes from MIMIC-III[19]. For the ShARe/CLEF and TAC2017ADR datasets, all the three BioBERT models outperformed the BERT$_{Base\_Cased}$ model and both the two ClinicalBERT models outperformed the corresponding BERT$_{Base\_Cased}$ and BERT$_{Large\_Cased}$ models. For the NCBI dataset, BioBERT$_{Base\_Cased+PubMed+PMC}$ and ClinicalBERT$_{Base\_Cased+MIMIC}$ were better than BERT$_{Base\_Cased}$. These indicate that the domain-specific BioBERT and ClinicalBERT are more appropriate than BERT for biomedical entity normalization. It would be interesting to pre-train a new bidirectional language representation model from scratch (or initialized with BERT$_{Base}$ or BERT$_{Large}$) using a large amount of drug labels from dailymed[37] and evaluate their effects on the TAC2017ADR dataset. We plan to conduct these studies in future.

The best performance was achieved when fine-tuning BioBERT$_{Base\_Cased+PubMed}$ for both the ShARe/CLEF and TAC2017ADR datasets, and when fine-tuning BioBERT$_{Base\_Cased+PubMed+PMC}$ for the NCBI dataset. This indicates that the model (i.e., BioBERT$_{Base\_Cased+PubMed+PMC}$) initialized with BERT$_{Base\_Cased}$ and pre-trained with both PubMed abstracts and PubMed Central full-text articles is effective for the NCBI dataset, and the pre-trained model (i.e., BioBERT$_{Base\_Cased+PubMed}$) with only PubMed abstracts is useful for both the ShARe/CLEF and TAC2017ADR datasets as well. This also illustrates that PubMed Central full-text articles are helpful for the PubMed abstracts but not for the clinical text and drug labels.

From Table 3, we notice that our best fine-tuned BERT-based ranking consistently outperformed the CNN-based ranking on all the three datasets, which indicates that pre-trained contextualized word representation models using bidirectional Transformers are more effective than the traditional context-independent word embeddings for the entity normalization task. Although the best fine-tuned models consistently outperformed previous state-of-the-art methods on all the three datasets, the improvements on the ShARe/CLEF and NCBI datasets were 0.35% and 0.26%, which was less than that on the TAC2017ADR dataset (i.e., 1.17%). For the ShARe/CLEF dataset, the main reason may be that we may not have completely reconstructed the ontology used in previous work[2–4], which was not released. For the NCBI dataset, the best performance was from TaggerOne (i.e., 88.80) which was reported by Leaman and Lu[36]. Their model was a joint model, which performed named entity recognition (with gold entity mentions as input) and normalization simultaneously. Such joint models could often leverage more contextual information to achieve better performance[36,38]. In the future, we will also investigate joint models to further improve entity normalization performance.

At this time, we applied and evaluated the pre-trained BERT / BioBERT / ClinicalBERT models for candidate concept ranking by transforming the ranking task as a sentence-pair classification task, which was a pointwise learning to rank method. We will further investigate pairwise learning to rank methods as used in previous work[6,7]. We are also planning to introduce the features used in Xu et al.'s system[6] into the final classifier layer of the candidate concept ranking module.

**Conclusion**

In this study, we applied and evaluated pre-trained language representation models for entity normalization using three biomedical datasets of different types. Preliminary results show that fine-tuning the pre-trained language representation models effectively advanced the state-of-the-art for biomedical named entity normalization.

**Acknowledgement**

This work is supported by NLM 5R01LM010681, NCI U24 CA194215, and NIGMS 5U01TR002062. Part of this work is supported by NVIDIA Corporation with the donation of the Quadro P6000 GPU.

**Conflicts of Interest**

Dr. Xu and The University of Texas Health Science Center at Houston have research-related financial interests in Melax Technologies, Inc.